\documentstyle[11pt,epsf,newpasp,twoside]{article}
\newcommand{\kms}{$\mbox{km~s}^{-1}$}
\newcommand{\myr}{$\mbox{mas~yr}^{-1}$}

\def\edcomment#1{\iffalse\marginpar{\raggedright\sl#1\/}\else\relax\fi}
\reversemarginpar

\begin{document}
\title{The Vela pulsar, the key?}
\author{Richard Dodson}
\affil{ISAS, Sagamihara, Kanagawa, Japan, 229-8510}
\author{Dion Lewis, David Legge, Peter McCulloch; John Reynolds, David
McConnell; Avinash Deshpande}
\affil{University of Tasmania, Hobart; ATNF, CSIRO, Australia; NAIC, PR}

\begin{abstract}
Of all pulsars known Vela has been one of the most productive in terms
in understanding pulsars and their characteristics. We present the
latest results derived from Australian telescopes. These include a
more accurate pulsar distance, a more precise pulsar local space
velocity, a new model of the spin up and the association of a radio
nebula with the X-ray pulsar wind nebula.
\end{abstract}

\vspace{-0.5cm}
\section*{Introduction}
We have observed the pulsar Vela with a range of Australian
telescopes. Using the Long Baseline Array (LBA) we have measured the
parallax of pulsar, and thus the distance. Using the Australia
Telescope Compact Array (ATCA) we have found the radio pulsar wind
nebula (PWN) that surrounds the X-ray PWN. Using the University of
Tasmania's dedicated pulsar monitoring telescope in Hobart we have
detected the core interaction in the spin up of the pulsar in the
glitch of 2000. 

\section*{Observations}
The pulsar monitoring telescope at Hobart is a fourteen meter radio
telescope dedicated to timing the Vela pulsar. It collects three
frequencies (635, 990, 1340-MHz), and the central one is collected
unfolded for high resolution timing analysis (\cite{dodson_glitch}). The
`fast component' observed, which was fitted with a decay time 1.2
minutes, has been reanalysed with a more realistic model. We marginally
detect the core interaction in the spin up (\cite{lewis_phd}). 

The ATCA has been used to map the radio PWN at 21cm, 13cm, 6cm and
3cm. Because we used compact configurations with better sensitivity to
low surface brightness objects we are able to map the whole nebula,
unlike previous observations (\cite{biet}).

We have used a single baseline from the LBA to measure the on sky
motion of the Vela pulsar compared with the extra galactic source
Vela-G. We have measured the proper motion and parallax of the Vela
pulsar to an unprecedented accuracy ($\mu_{\alpha {\rm cos}\delta}=
-49.68 \pm 0.06,\ \mu_\delta= 29.9 \pm 0.1$ \myr\ , $\pi = 3.5 \pm
0.2$ mas), and have been able to convert these back to the space
velocity and position angle of the pulsar in its local environment
with greater precision that previously possible ($61 \pm 2$ \kms at
$301^\circ\pm1.8$), because of the unambiguity in the radio reference
frame. We have found an arithmetic error in \cite{caraveo_pm} and,
once corrected, their results agree with ours (Caraveo, personal
comms).

\section*{Future observations}

We have funding for a coherently dedispersed 30MHz backend for the
635MHz IF on the pulsar timing telescope. This should allow us an
increase in sensitivity of an order of magnitude over the previous
observations.

We plan to observe at the ATCA the radio nebula at higher frequencies
to find the turn over frequency, and model the emission from the
X-rays down to the radio frequencies. We are observing at the VLA to
get sensitive rapid observations to measure changes associated with
the recently discovered X-ray outer jet (\cite{pavlov_vari}).

The limitation in the accuracy of the VLBI observation is the solar
motion parameters, and we can not improve on this. Nevertheless we
are planning to use Vela as a demonstration source in a baseband
e-VLBI experiment.

\begin{figure}
\plottwo{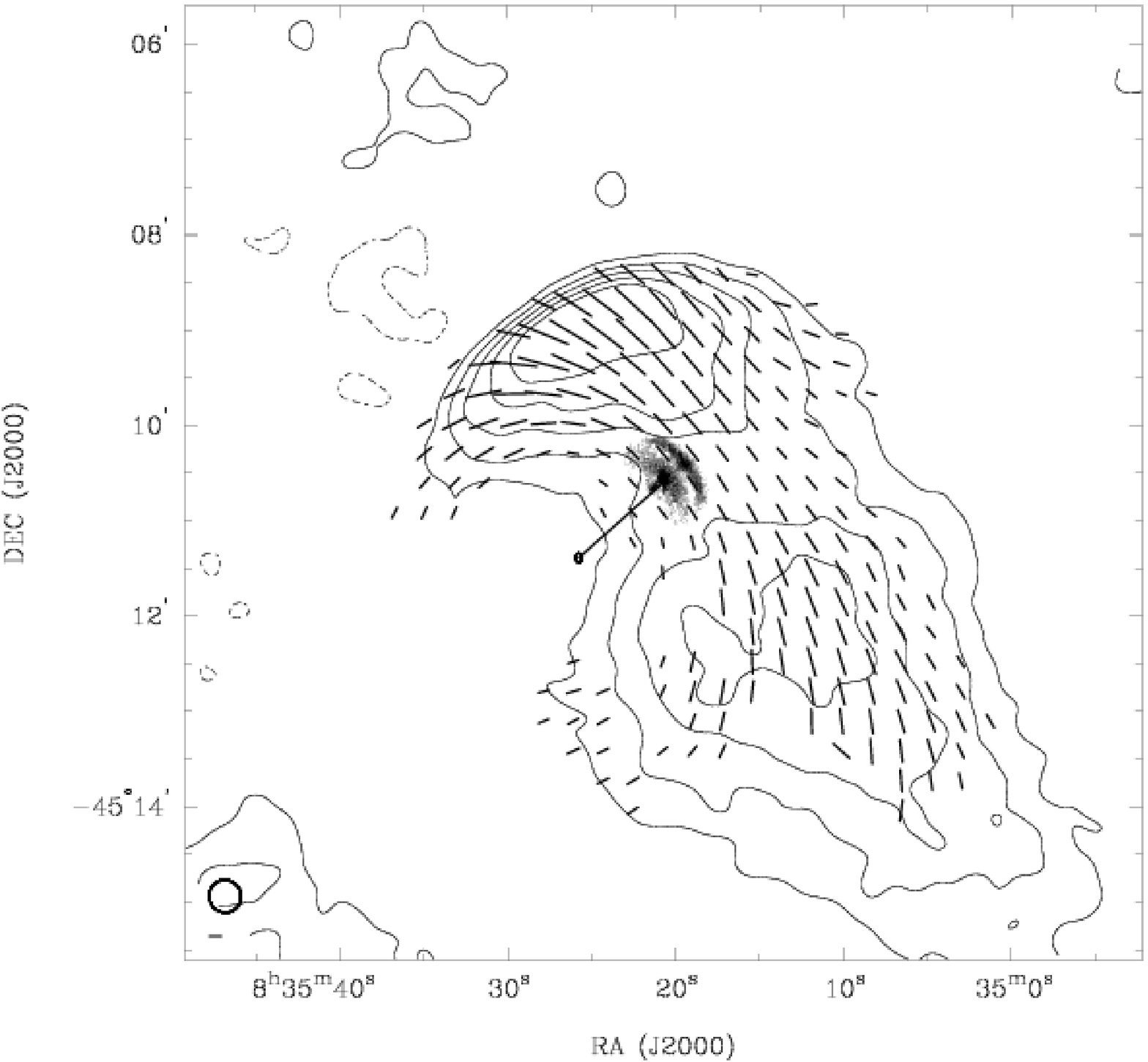}{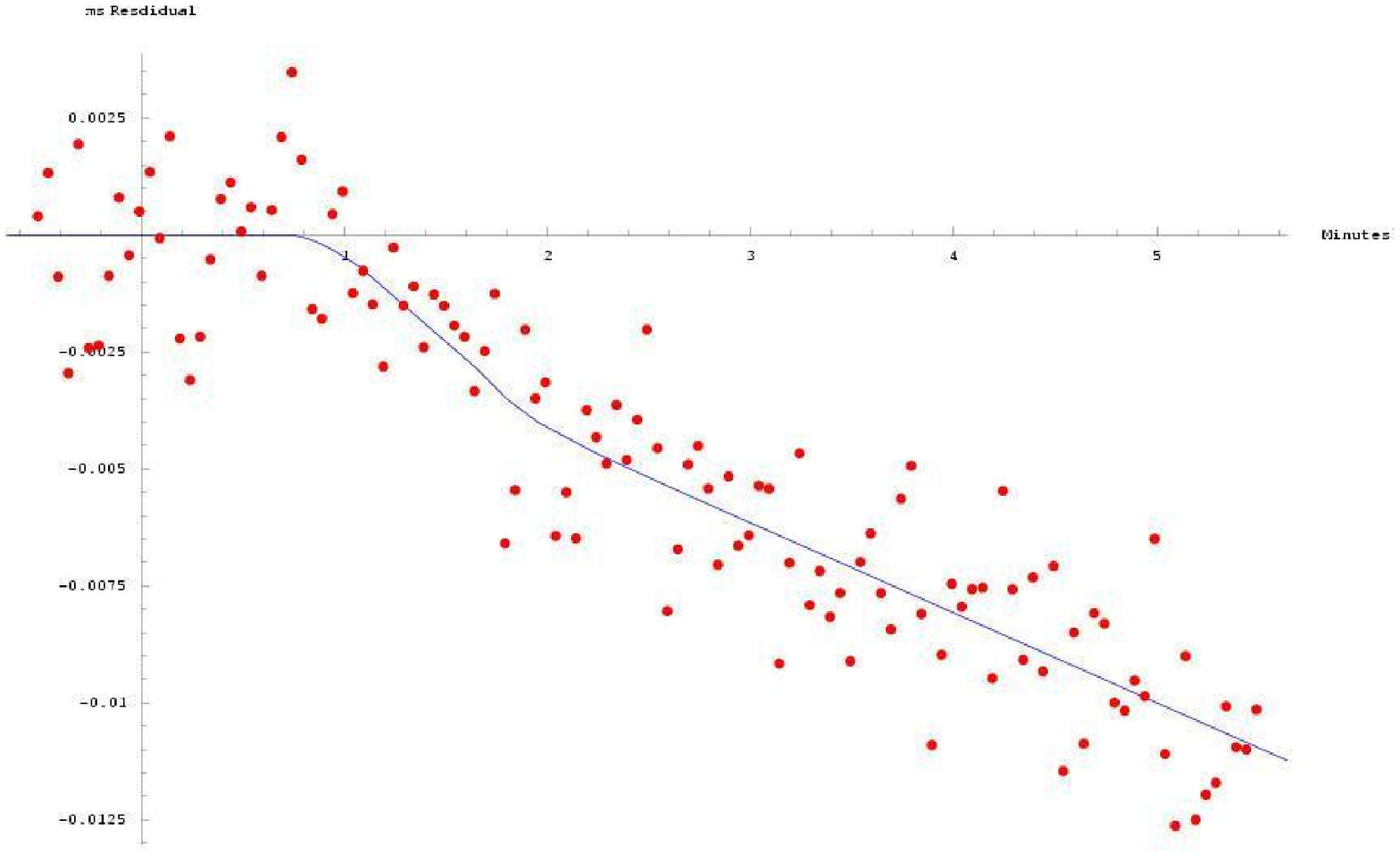}
\caption{a) The Chandra observation of the Vela PWN (grey scale) and
the 5~GHz Radio contours (-1,1,2,3,4,5 mJy/beam). The derotated
magnetic field lines are overlaid. The proper motion vector shows
the distance travelled in 1000 years, and ends with the three sigma
error ellipse. b) 3 second integrations of the single pulse data
across the 2000 glitch. The fit is for the new model of crust-core
interaction, and illustrates the limits on this from the data.}
\end{figure}

\begin{footnotesize}

\end{footnotesize}
\end{document}